\documentclass{ifacconf}

\usepackage{graphicx}      
\usepackage{natbib}        

\usepackage{url}
\usepackage{array,tabularx}
\usepackage{multicol} 
\usepackage{multirow} 
\usepackage{amssymb}
\usepackage{amstext}
\usepackage{amsmath}
\usepackage{mathtools}
\usepackage{mathrsfs}
\usepackage{xcolor}
\usepackage{graphicx}
\usepackage{subcaption}
\usepackage{epsfig}
\usepackage{makecell}

\begin{document}
\begin{frontmatter}
\textcolor{blue}{This work has been submitted to IFAC for possible publication.}
\title{A Quantum Algorithm for the Diffusion Step of Grid-based Filter\thanksref{footnoteinfo}} 

\thanks[footnoteinfo]{This work was supported in part by the National Research Foundation of Korea(NRF) grant 
funded by the Korea government(MSIT)(RS-2025-24535123, RS-2025-00462298) and in part by the Czech Science Foundation (GACR) under grant GA 25-16919J.}

\author[First]{Yeongkwon Choe} 
\author[Second]{Chan Gook Park}
\author[Third]{Jind\v{r}ich Dun\'ik, Jan Krejčí}
\author[Fourth]{Jakub Matou\v{s}ek} 
\author[Fifth]{Marek Brandner}

\address[First]{Dept. of Mechatronics Eng., Kangwon National Univ.
      Chuncheon, Republic of Korea (e-mail: ychoe@kangwon.ac.kr)}
\address[Second]{Dept. of Aerospace Eng./ASRI, Seoul National Univ.,
      Seoul, Republic of Korea (e-mail: chanpark@snu.ac.kr)}
\address[Third]{Dept. of Cybernetics, Univ. of West Bohemia in Pilsen, Czech Republic (e-mail: \{dunikj, jkrejci\}@kky.zcu.cz)}
\address[Fourth]{Oden Institute for Computational Engineering and Sciences, The University of Texas at Austin, TX, USA (e-mail: jakub.matousek@austin.utexas.edu)}
\address[Fifth]{Dept. of Mathematics, Univ. of West Bohemia in Pilsen, Czech Republic (e-mail: brandner@kma.zcu.cz).}

\begin{abstract}                
We propose a simple quantum algorithm for implementing the diffusion step of grid-based Bayesian filters. 
The method encodes the advected state density and the process noise density into quantum registers and realizes diffusion using a quantum Fourier transform--based adder. 
This avoids the explicit convolution required in classical implementations and the repeated coin-flip operations used in quantum random walk approaches. 
Numerical simulations using a gate-based quantum computing simulator confirm that the proposed approach reproduces the desired probability densities while requiring significantly fewer quantum gates and much shallower circuit depth.
\end{abstract}


\end{frontmatter}

\section{Introduction}
With the rapid development of quantum computing technologies, 
research on quantum algorithms is becoming increasingly important across many areas of science and engineering. 
A key characteristic of quantum computation is that a quantum state can exist in a superposition of many basis states, 
allowing a quantum processor to represent and manipulate multiple possibilities simultaneously. 
Moreover, the tensor-product structure of multi-qubit systems provides considerably large representation space, 
enabling more efficient processing for high-dimensional systems, which is often difficult using classical methods. 
In particular, search and optimization problems have been a natural focus of research in these areas.
More recently, the ability to operate in extremely high-dimensional feature spaces has drawn growing interest in machine learning, 
making quantum approaches a promising direction for addressing challenges encountered in modern data-driven applications (\cite{KuSaUy:21}).

In the field of state estimation, the application of quantum computing has only recently begun to attract attention, 
and relevant studies remain in an early stage.\footnote{
    The term \textit{quantum estimation} typically refers to estimating an unknown quantum state.
    Here, however, we use quantum computing to address sequential estimation of a state vector corresponding to a dynamical system, which is a distinct problem.
}
For example, \cite{NiMa:19} demonstrated the feasibility of simulating Markov process transitions on real quantum hardware.
\cite{Po:23} described quantum versions of several Bayesian computational methods, including MCMC (Markov chain Monte Carlo), Gaussian processes, and high-dimensional regression, suggesting potential benefits for high-dimensional inference.
Around the same time, \cite{LaMa:23} proposed a quantum MCMC algorithm that is robust to noise on error-prone quantum processors and experimentally showed that it can achieve a cubic-to-quartic speedup over classical approaches.
In multitarget tracking, \cite{McKl:22} employed quantum annealing to address the combinatorial complexity of multitarget data association. 
Rather than relying on adiabatic quantum computing (AQC) to find a single optimal assignment, they intentionally used diabatic quantum annealers (DQA) to generate multiple high-probability (low-energy) feasible assignments, 
which were then combined with classical Bayesian filtering methods.
\cite{StUl:23} presented quantum optimization algorithms that are broadly applicable to multitarget data association and weapon target assignment problems,
including an AQC-based solver and a quantum approximate optimization algorithm (QAOA)–based solver that can be executed on gate-based quantum computers.

While quantum approaches have begun to show potential in accelerating subproblems in state estimation problems, 
the classical Bayesian filter itself remains at a very early stage of exploration, 
with many promising research directions still open for future developments.

\subsection{Bayesian Estimation using Grid-based Filter}
Bayesian filters are built upon Bayesian recursive relations (BRRs), which consist of a filtering step followed by a prediction step.
The filtering step calculates the posterior density using the Bayes’ rule, which can be written as
\begin{equation} \label{eq:filt}
      p(\mathbf{x}_k|\mathbf{z}^k)=\frac{p(\mathbf{x}_k|\mathbf{z}^{k-1})p(\mathbf{z}_k|\mathbf{x}_k)}{p(\mathbf{z}_k|\mathbf{z}^{k-1})},
\end{equation}
where $\mathbf{x}_k$ denotes the state at time $k$, $\mathbf{z}_k$ is the measurement at time $k$, 
$\mathbf{z}^k=\{\mathbf{z}_0,\ldots,\mathbf{z}_k\}$ is the measurement history, and $p(\mathbf{z}_k|\mathbf{x}_k)$ is the measurement model.
The prediction step then propagates this posterior forward in time through the process model $ p(\mathbf{x}_{k+1}|\mathbf{x}_{k})$ to get the prior density, as expressed by the Chapman–Kolmogorov equation (CKE)
\begin{equation} \label{eq:pred}
    p(\mathbf{x}_{k+1}|\mathbf{z}^{k})=\int p(\mathbf{x}_{k+1}|\mathbf{x}_{k})p(\mathbf{x}_{k}|\mathbf{z}^{k})d\mathbf{x}_{k}.
\end{equation}
For certain restricted class of models such as linear Gaussian systems, analytic solutions such as the Kalman filter are available.
However, in many practical scenarios, closed-form solutions to \eqref{eq:filt} and \eqref{eq:pred} are intractable; 
therefore, considerable research has focused on developing approximate methods for BRRs.

Grid-based filters(GbFs) constitute one such class of methods: 
they approximate the BRRs by partitioning the state space into a deterministic grid and 
evaluating the prediction and update steps directly on this discretized domain.
Although GbFs have been studied extensively, they still suffer from a high computational cost that grows exponentially with the dimension of the state space.
 Various strategies—such as functional decomposition (\cite{StDuTi:23, TiStDu:23}), 
copula-based decomposition (\cite{DuStMaBl:22, AjSt:23}), 
and tensor decomposition (\cite{MaBrDuPu:24,Go:18})—are still being actively explored to alleviate this burden.
Notably, the recent work of \cite{MaDuSt:25} revisits the discrete Fourier transform--based approach of \cite{PaZh:11} 
and recasts it within their Lagrangian framework. 
Their method reduces the computational complexity of the diffusion in the prediction step---the most computationally demanding part---from $O(N^2)$ to $O(N \log_2 N)$, 
where $N$ denotes the total number of grid points, a quantity that grows exponentially with the state dimension.
However, despite these ongoing efforts, it remains hard for classical computers 
to avoid the fundamental limitation of having to store and process each grid point individually. 

\subsection{Quantum State Estimation Algorithms: An Overview}
The fundamental unit of quantum information, the qubit, can exist in a superposition of two basis states. 
Thus, whereas a classical computer requires two bits to store two distinct states, a single qubit can represent both states simultaneously. 
Moreover, the multi qubit systems provide an exponentially large representation space:
for example, while a classical computer needs on the order of $10^8$ storage units 
and operations to represent a four-dimensional grid with 100 grid points per dimension, 
a quantum computer requires only 27 qubits to encode the entire state space (\cite{Go:24a}).
In addition, because the amplitudes of a qubit encode the measurement probabilities of its basis states, 
such quantum states are naturally suited for the representation of probability densities.

Two recent works (\cite{Go:24a, Go:24b}) propose quantum algorithms for the BRRs. 
As claimed by the author, the algorithm developed for the filtering step constitutes the first quantum algorithm for the Bayesian update. 
Both works rely on the \textit{quantum random walks} (QRW)(\cite{Ke:03}), with one addressing the prediction step 
and the other extending the approach to the filtering step. 
The method implements diffusion by representing a \textit{coin} with a single qubit and shifting the system state by $+1$ or $-1$ according to the outcome of a coin flip, a process often described as a simple symmetric random walk. 
The filtering method builds on the Daum--Huang particle flow filter (PFF), 
which reformulates the Bayesian update as an imaginary-time stochastic process (\cite{DaHuNo:10}); 
the quantum algorithm then implements this process using QRW.

Although these works provide an important first step toward extending Bayesian filtering into the quantum domain 
and their coin-flip-based incremental diffusion appears well suited for implementing continuous processes such as those in PFF, 
it may be inefficient for realizing the diffusion of discrete-time state--space models.
In such cases, many repeated QRW steps are required, which increases the quantum circuit depth and the number of qubits for additional coin flips. 
This can be a limitation, since noisy intermediate-scale quantum (NISQ) devices have so far been limited to relatively shallow circuit depth and are constrained in the number of available qubits.

\subsection{Goal of the Paper: Quantum Realized Diffusion GbF}
In this work, we show that diffusion can be implemented in a remarkably simple manner by leveraging the quantum Fourier transform (QFT), enabling the entire operation to be executed in a single step. The main idea is that, similar to representing a system state with an quantum register, an additional quantum register can be allocated to represent the system noise. Diffusion is then carried out simply by adding the noise register to the state register according to the system model. In classical computation, this step is computationally demanding because addition in the state space corresponds to convolution in the probability density. Surprisingly, quantum computation offers a more direct mechanism: by encoding probability values into the probability amplitudes of qubits, the measurement statistics of two added quantum states naturally correspond to the convolution of their associated probability densities. Consequently, the diffusion step reduces to a simple register addition.

The evolution of a closed quantum system preserves information, requiring all quantum operations to be unitary and hence reversible. However, classical adders are inherently irreversible because their outputs do not uniquely determine the original inputs; Implementing a classical adder on quantum hardware would therefore require additional ancilla qubits, creating unnecessary overhead. Instead, we adopt a QFT-based quantum adder, which is specifically designed for addition in quantum computation (\cite{Dr:00,RuGaJc:17}). Our analysis shows that this approach reproduces the diffusion step with reduced complexity compared to conventional algorithms. 

The remainder of this paper is organized as follows.
Section \ref{sec:2} provides a brief overview of GbFs, and Section \ref{sec:3} presents a simple QFT-based algorithm for implementing the diffusion step in GbFs. 
Section \ref{sec:4} demonstrates the presented method and compares its computational complexity with that of the conventional diffusion algorithm.
Finally, Section \ref{sec:5} concludes this work.

\section{Preliminaries}
\label{sec:2}
\subsection{State-Space Model}
This work considers the discrete-time state space model with additive noise, which is typically assumed in GbF, as follows:
\begin{align}
      \mathbf{x}_{k+1} &= \mathbf{f}_k(\mathbf{x}_{k},\mathbf{u}_{k})+\mathbf{w}_k ,  \label{eq:ssmx}\\
      \mathbf{z}_{k} &= \mathbf{h}_k(\mathbf{x}_{k})+\mathbf{v}_k, \label{eq:ssmz}
\end{align}
Here, $\mathbf{x}_{k} \in\mathbb{R}^{n_x}$, $\mathbf{u}_k \in \mathbb{R}^{n_u}$, and $\mathbf{z}_k \in \mathbb{R}^{n_z}$ denote 
the system state, the control input, and the measurement at time $k$, respectively.
The process noise $\mathbf{w}_k$ and measurement noise $\mathbf{v}_k$ are assumed to be mutually independent with known probability density functions (PDFs), 
and the functions $\mathbf{f}_k(\cdot)$ and $\mathbf{h}_k(\cdot)$ represent the state transition and measurement models, respectively.

\subsection{Grid-based Filtering and Problem Statement}
GbF approximately solves the BRRs in \eqref{eq:filt}–\eqref{eq:pred} for the model 
\eqref{eq:ssmx}–\eqref{eq:ssmz} by using a piecewise constant point-mass approximation of the posterior:
\begin{equation}
      \label{eq:pmd}
      p(\mathbf{x}_k\mid\mathbf{z}^{k})\approx \hat{p}(\mathbf{x}_k\mid\mathbf{z}^{k};\Xi_k)
      \triangleq\sum_{i=1}^{N}\hat{p}(\xi_k^{(i)}\mid\mathbf{z}^{k})\mathbf{1}_{\Psi_k^{(i)}}(\mathbf{x_k})
\end{equation}
where $\Xi_k = \{\xi_k^{(i)}\in\mathbb{R}^{n_x}\}_{i=1}^{N}$ denotes the set of grid points, 
and $\hat{p}(\xi_k^{(i)}\mid\mathbf{z}^{k})$ represents the approximate probability mass for the grid point $\xi_k^{(i)}$.
The indicator function $\mathbf{1}_{\Psi_k^{(i)}}(\mathbf{x}_k)$ selects
the hyper-rectangular cell $\Psi_k^{(i)} \subset \mathbb{R}^{n_x}$ surrounding $\xi_k^{(i)}$, defined as
\begin{equation}
      \begin{aligned}
            \Psi_k^{(i)}
            =
            \bigl[\xi_k^{(i)}(1) - \Delta_k(1)/2,\; \xi_k^{(i)}(1) + \Delta_k(1)/2\bigr]
            \times \cdots \\ \times
            \bigl[\xi_k^{(i)}(n_x) - \Delta_k(n_x)/2,\; \xi_k^{(i)}(n_x) + \Delta_k(n_x)/2\bigr],
      \end{aligned}
\end{equation}
where $\xi_k^{(i)}(j)$ denotes the $j$-th component of $\xi_k^{(i)}$, 
and $\Delta_k(j)$ is the grid spacing along the $j$-th state dimension.

In this work, we adopt a Lagrangian perspective for the prediction step of the GbF, instead of the classical Eulerian perspective (\cite{MaDuSt:25}). 
Under the Lagrangian approach, computing the prediction density in \eqref{eq:pred} for the state transition model \eqref{eq:ssmx} 
is decomposed into an \textit{advection} part, which evaluates the flow induced by the dynamics, 
and a \textit{diffusion} part, which accounts for the process noise. 
This decomposition transforms the most computationally demanding component of the GbFs from a generalized convolution into a standard convolution, thereby enabling an efficient solution using the DFT.

In particular, the advection step for~\eqref{eq:ssmx} approximates the PDF of the state transformed by the transition function, i.e., of
\begin{equation}
    \label{eq:adv}
    \mathbf{x}_{k+1}^{\mathrm{adv}} 
    = \mathbf{f}_k(\mathbf{x}_k, \mathbf{u}_k),
\end{equation}
and each grid point $\xi_k^{(i)}$ is transported according to this mapping. 
Interpolation is carried out either before or after advection to realign the advected density with the regularly spaced grid $\Xi_k^{\mathrm{adv}}$ designed to support the advected PDF, and to incorporate measurement information and dynamics uncertainty into the grid design.
The diffusion step starts from 
\begin{equation}
    \label{eq:dif_prev}
    \mathbf{x}_{k+1} 
    = \mathbf{x}_{k+1}^{\mathrm{adv}} + \mathbf{w}_k,
\end{equation}
and calculates the prior density ${p}(\mathbf{x}_{k+1}\mid\mathbf{z}^{k})$ using a convolution of $p(\mathbf{x}_{k+1}^{\mathrm{adv}}\mid\mathbf{z}^{k})$, resulting from the advection \eqref{eq:adv} 
and the process noise density $p(\mathbf{w}_k)$, as
\begin{equation}\label{eq:dif}
    {p}(\mathbf{x}_{k+1}\mid\mathbf{z}^{k})=\int p(\mathbf{x}_{k+1}^{\mathrm{adv}}\mid\mathbf{z}^{k})p_{\mathbf{w}_k}(\mathbf{x}_{k+1}-\mathbf{x}_{k+1}^{\mathrm{adv}})d\mathbf{x}_{k+1}^{\mathrm{adv}}.
\end{equation}
Numerical solution to \eqref{eq:dif} on a grid $\Xi_k$ using the DFT leads to the point-mass approximation of the prior $\hat{p}(\mathbf{x}_{k+1}\mid\mathbf{z}^{k};\Xi_{k+1})$.

In this work, rather than evaluating this convolution (as the most complex part of the GbF) explicitly in the classical computing manner using the DFT,
we encode the probability densities into quantum amplitudes and realize \eqref{eq:dif} 
as an addition between superposed quantum states.
The advection and measurement update steps, for which quantum advantages are not particularly expected, 
are assumed to be performed on a classical processor.
Further details of GbFs, including the measurement update, can be found in the existing literature (\cite{MaDuSt:25}).

\section{QFT-based Simple Quantum Diffusion}
\label{sec:3}
\subsection{Qubits and Their Representation} 
Similar to a classical bit, a \textit{qubit} is the fundamental unit of quantum information and encodes a two-state quantum system.
A convenient choice of orthonormal basis for this space is $\{|0\rangle, |1\rangle\}$, expressed using Dirac notation\footnote{Dirac (bra--ket) notation is the standard representation of quantum states. 
A state $|\psi\rangle$ is written as a \textit{ket}, and its conjugate transpose $\langle\psi|$ as a \textit{bra}. 
Inner and outer products are expressed as $\langle\phi|\psi\rangle$ and $|\psi\rangle\langle\phi|$.}. 
A general single qubit state can be written as a \textit{superposition} of these basis states:
\begin{equation}
    |\psi\rangle = \alpha |0\rangle + \beta |1\rangle,
\end{equation}
where $\alpha, \beta \in \mathbb{C}$ and $|\alpha|^2 + |\beta|^2 = 1$, and are referred to as the \textit{amplitudes}.
The state $|\psi\rangle$ exists as a superposition of the basis states, 
and upon measurement it collapses to either $|0\rangle$ or $|1\rangle$ with probabilities $|\alpha|^2$ and $|\beta|^2$, respectively. 

A multi qubit system is described by the tensor (Kronecker) product of individual qubit states. 
For example, the state of two qubits $|\psi_1\rangle$ and $|\psi_2\rangle$ is written as
\begin{equation}
    |\psi_1\rangle \otimes |\psi_2\rangle.
\end{equation}
A general $n$-qubit system represented in a $2^n$-dimensional Hilbert space, 
and its computational basis consists of tensor products of single qubit basis states, 
such as $|b_{n-1}\rangle \otimes \cdots \otimes |b_0\rangle$ with $b_j \in \{0,1\}$.
This tensor-product structure allows a quantum processor to represent $2^n$ basis states simultaneously and forms the basis for quantum parallelism.

\subsection{Quantum Encoding of State and Noise Grids.}
To implement the diffusion step \eqref{eq:dif} on a quantum processor, 
the process noise is also represented by a point-mass approximation using a noise grid 
$\Omega_k = \{\omega_k^{(i)} \in \mathbb{R}^{n_x}\}_{i=1}^{N}$ 
that has the same spacing $\Delta_k(j)$ as the state grid. 
Both the state grid and the noise grid are embedded into quantum registers through 
affine (translation–scaling) mappings that convert each dimension of the continuous state space 
into an integer index compatible with the computational basis of multi qubit registers.
Because the computational basis encodes nonnegative integers by default, while process noise is commonly modeled as zero-mean, 
we apply two’s complement encoding so that the resulting indices can be represented as signed integers.

For each dimension $j$, the state grid points $\Xi(j)$ along that coordinate have uniform spacing 
$\Delta(j)$ and its lower bound $\xi_{\min}(j)$.
This allows us to define the one-dimensional affine index map
\begin{equation}
      \label{eq:tj_def}
      \begin{aligned}
            &T_j :\; \{\xi^{(i)}(j) \in \Xi(j)\}_{i=1}^{n_j}
                  \rightarrow \{0,1,\dots,2^{n_j}-1\},\\
            &T_j\bigl(\xi(j)\bigr) = \frac{\xi(j) - \xi_{\min}(j)}{\Delta(j)},
      \end{aligned}
\end{equation}
where $n_j$ denotes the number of grid points along $j$-th dimension, such that $N=\prod\nolimits_{j=1}^{n_x}n_j$.
For the advected grid point $\xi_k^{\mathrm{adv},(i)} $,  
its quantum representation is the tensor-product computational basis state
\begin{equation}
    |\ell^{(i)}_{k+1}\rangle = \bigotimes_{j=1}^{n_x}|\ell_{k+1}^{(i)}(j)\rangle,
\end{equation}
where
\begin{equation}
    \ell_{k+1}^{(i)}(j) = T_j\left( \xi_{k+1}^{\mathrm{adv},(i)}(j) \right).
\end{equation}

The advected PDF over the grid $\hat{p}(\mathbf{x}_{k+1}^{\mathrm{adv}}\mid\mathbf{z}^{k};\Xi_k^{\mathrm{adv}})$ is then encoded as
\begin{equation}
    |\psi_{k+1}^{\mathrm{adv}}\rangle =
    \sum_{i=1}^{N} \sqrt{\hat{p}(\mathbf{x}_{k+1}^{\mathrm{adv}}=\xi_{k+1}^{\mathrm{adv},(i)}\mid\mathbf{z}^{k})}|\ell^{(i)}_{k+1}\rangle.
    \label{eq:x_encod}
\end{equation}

For the process noise grid $\Omega_k$, each dimension is constructed using the same number 
of unique grid coordinate values $n_j$ and the same spacing $\Delta_j$ as the corresponding 
state grid. To maintain this spacing and indexing structure, we first form the integer indices 
associated with each coordinate and then assign the corresponding noise grid points 
$\omega_k^{(i)}(j)$ to these indices.
Since the process noise is typically modeled as zero-mean, it's representation requires both positive and negative values.
We thus employ \emph{two's-complement} encoding (see below) to represent signed integers in the quantum registers. This choice is mainly for implementation convenience and ensures a uniform treatment of signed quantities.

We define the signed integer corresponding to $\omega(j)$ as
\begin{equation}
    S_j\bigl(\omega(j)\bigr) \in \{-2^{n_j-1},\dots,2^{\,n_j-1}-1\}
\end{equation}
and express the corresponding grid point as
\begin{equation}
    \omega(j) = S_j\bigl(\omega(j)\bigr)\Delta_j.
    \label{eq:sj_def}
\end{equation}
The two's-complement (unsigned) index stored in the quantum register is then
\begin{equation}
\tilde{S}_j\bigl(\omega(j)\bigr)
=
\left( S_j\bigl(\omega(j)\bigr) \bmod 2^{n_j} \right)
\in \{0,\dots,2^{n_j}-1\}.
\end{equation}

Each noise grid point $\omega_k^{(i)} \in \mathbb{R}^{n_x}$ is mapped to the tensor-product 
computational basis state
\begin{equation}
    |m_k^{(i)}\rangle = \bigotimes_{j=1}^{n_x}\bigl|\tilde{S}_j\!\left(\omega_{k}^{(i)}(j)\right)\bigr\rangle.
\end{equation}

The noise PDF is encoded into the amplitude of the quantum register as
\begin{equation}
    |\phi_k\rangle =\sum_{i=1}^N \sqrt{p(\mathbf{w}_k=\omega_k^{(i)})}|m_k^{(i)}\rangle.
    \label{eq:w_encod}
\end{equation}

\subsection{Quantum Realization of the Diffusion Step}
In our construction, each dimension $j$ of the state space is assigned a dedicated multi qubit subregister, 
enabling the diffusion step \eqref{eq:dif} to be carried out dimension-wise using a QFT-based quantum adder.
Although the global quantum states $|\psi_{k+1}^{\mathrm{adv}}\rangle$ and $|\phi_k\rangle$ may in general be entangled across their coordinate subregisters, 
the underlying Hilbert space remains a tensor product of the dimension-wise subspaces.
Consequently, the adder for dimension $j$ is implemented as a unitary acting only on the $j$-th subregister, $U_\mathrm{add}^j$, leaving the remaining subregisters unchanged.
The action of $U_{\mathrm{add}}^{(j)}$ on the $j$-th subregister is
\begin{align}
    &|\tilde{S}_j(\omega_k^{(i)}(j))\rangle |\ell_{k+1}^{(i)}(j)\rangle
     \xmapsto{\,U_{\mathrm{add}}^{(j)}\,}
    \nonumber\\
    &\qquad
    |\tilde{S}_j(\omega_k^{(i)}(j))\rangle
    \bigl|\ell_{k+1}^{(i)}(j) + \tilde{S}_j(\omega_k^{(i)}(j)) 
      \pmod{2^{n_j}}\bigr\rangle , 
    \label{eq:add_local}
\end{align}
with the identity acting on all other dimensions, so the full operation is $U_{\mathrm{add}}^{(j)} \otimes I_{\mathrm{rest}}$.
The amplitude of the resulting quantum state consequently represents the discrete convolution between the advected density and the process noise density.

To implement $U_{\mathrm{add}}^{(j)}$ on the $j$-th subregister, we employ the QFT-based quantum adder introduced by \cite{Dr:00}. 
For clarity of notation, the operation in \eqref{eq:add_local} can be rewritten in terms of 
two $n$-qubit registers encoding unsigned integers $a_u$ and $b_u$ as
\begin{equation}
    U_{\mathrm{add}}|a_u\rangle|b_u\rangle=|a_u\rangle|a_u+b_u\pmod {2^n}\rangle.
\end{equation}
In the context of the diffusion step, one may interpret $a_u$ as the process noise and $b_u$ as the advected state.
The addition unitary acting on these registers is expressed as
\begin{equation}
    U_{\mathrm{add}} = \left(I \otimes F_{2^n}^\dagger\right)
    \left(\sum_{a_u=0}^{2^n-1} |a_u\rangle\langle a_u| \otimes R(a_u)\right)
    \left(I \otimes F_{2^n}\right),
    \label{eq:draper_unitary}
\end{equation}
where $F_{2^n}$ denotes the QFT on $2^n$ basis states, its adjoint, $F_{2^n}^\dagger$, is the inverse QFT,
and $R(a_u)$ is the diagonal phase operator defined in the Fourier basis as
\begin{equation}
R(a_u)\,|k\rangle = e^{2\pi i a_u k/2^n}\,|k\rangle .
\end{equation}
In short, the adder operates by transforming the target register $|b\rangle$ into the Fourier basis via the QFT, 
applying a phase rotation determined by $a_u$, and then returning to the computational basis through the inverse QFT.
To aid understanding, $\sum |a_u\rangle\langle a_u|$ serves to condition the phase operator $R(a_u)$ on the control value $a_u$, 
ensuring that the appropriate phase rotation is applied for each basis state of $|a_u\rangle$.
Because the addition is realizaed as a phase shift, the operation naturally yields the \textit{modular} addition.

\subsection{Consistency Check: Two's-Complement Addition}
For completeness, we briefly check that modular addition between a signed integer register and a two’s-complement encoded register produces the correct signed result.
Let $a_s$ denote the original signed integer value and $a_u$ its two’s-complement encoding; we show that 
\begin{equation}
    (b_u + a_u)\bmod 2^n = (b_u + a_s)\bmod 2^n.
\end{equation}

\subsubsection*{Case 1: $a_s \ge 0$}
When $a_s \ge 0$, we have $a_u = a_s$, and therefore
\begin{equation}
(b_u + a_u) \bmod 2^n = (b_u + a_s) \bmod 2^n.
\end{equation}

\subsubsection*{Case 2: $a_s < 0$}
When $a_s < 0$, two's-complement encoding satisfies
\begin{equation}
a_u = 2^n - |a_s|.
\end{equation}
Then
\begin{align}
(b_u + a_u) \bmod 2^n &= (b_u + 2^n - |a_s|) \bmod 2^n \\\nonumber
&= (b_u - |a_s|) \bmod 2^n \\\nonumber
&= (b_u + a_s) \bmod 2^n.
\end{align}

\section{Numerical Implementation}
\label{sec:4}
\begin{figure*}[t]
    \centering
    \includegraphics[width=0.95\linewidth]{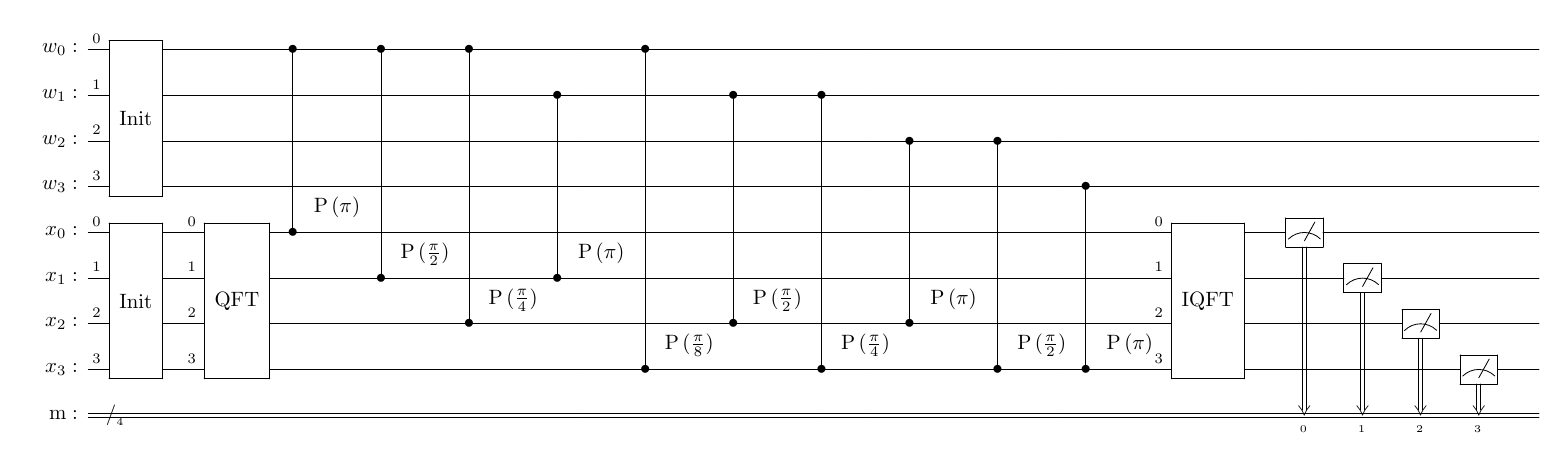}
    \vskip -0.5pc
    \caption{Quantum circuit for the proposed QFT-based diffusion.} 
    \label{fig:proposed_qc}
\end{figure*}

In this section, we implement the proposed quantum algorithm for the diffusion step and compare it against the existing QRW-based approach. 
Although the numerical results are straightforward, our aim is to confirm that the proposed method attains the desired diffusion result while using substantially fewer quantum resources. 
All implementations are performed using Qiskit 2.1, a quantum computing framework developed by IBM, 
and the simulation results are obtained with AerSimulator 0.17, which is included within the Qiskit environment.

Following the evaluation setting of (\cite{Go:24a}), we consider a one-dimensional state space and use four qubits to represent the state.
Figure \ref{fig:proposed_qc} shows the resulting quantum circuit for the proposed approach. We evaluate four test cases:
\begin{itemize}
    \item \textbf{Case 1:} $\mathbf{x}^{\mathrm{adv}} \sim \mathcal{N}(7,1^2)$, 
    $\mathbf{w} \sim \mathcal{N}(0,1^2)$; \\QRW: 1 repeat to approximate noise.

    \item \textbf{Case 2:} $\mathbf{x}^{\mathrm{adv}} \sim \mathcal{N}(7,1^2)$, 
        $\mathbf{w} \sim \mathcal{N}(0,2^2)$; \\QRW: 4 repeats to approximate noise.

    \item \textbf{Case 3:} $\mathbf{x}^{\mathrm{adv}} = 7$, 
        $\mathbf{w} \sim \mathcal{N}(0,1^2)$; \\QRW: 4 repeats to approximate noise.

    \item \textbf{Case 4:} $\mathbf{x}^{\mathrm{adv}} = 7$, 
        $p(\mathbf{w}) = \{1/16,\, 1/4,\, 3/8,\, 1/4,\, 1/16\}$ on $\{-4,-2,0,2,4\}$; QRW: 4 repeats.
\end{itemize}

In this work, the QRW-based approach is used to approximate the Gaussian distribution by adjusting the number of repeats according to $\mathrm{Var}(w)$, motivated by the fact that repeated quantum random walks converge to a Gaussian distribution via the central limit theorem. 
As a result, a small number of repeats inevitably leads to an inaccurate approximation.
Case 4 considers the opposite situation: rather than using a QRW-based approach to approximate a target distribution, we encode into the proposed QFT-based method a distribution that a QRW can generate exactly.
This demonstrates that the same distribution can be realized with substantially lower quantum-resource requirements using the proposed approach.

\begin{figure*}[t]
    \centering
    \begin{subfigure}{0.495\linewidth}
        \centering
        \includegraphics[width=1\linewidth]{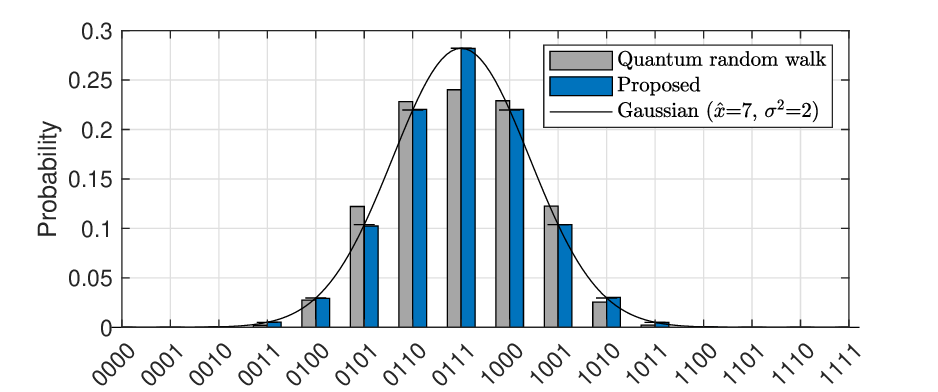}
        \vskip -0.5pc
        \caption{}
        \label{fig:guassinit_1step}
    \end{subfigure}
    \hfill
    \begin{subfigure}{0.495\linewidth}
        \centering
        \includegraphics[width=1\linewidth]{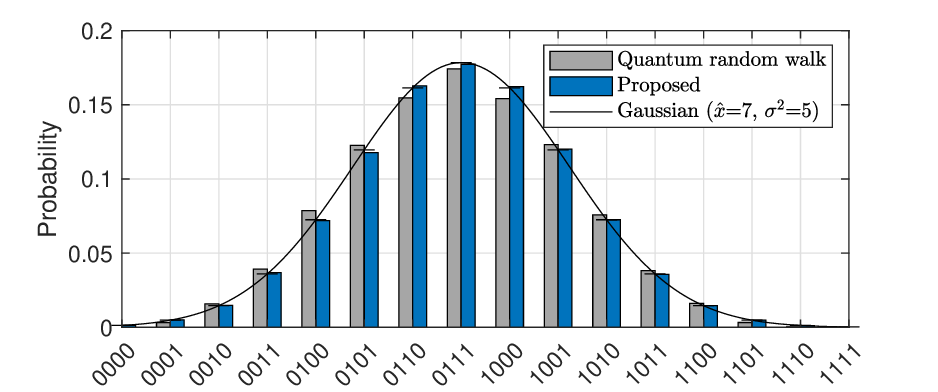}
        \vskip -0.5pc
        \caption{}
        \label{fig:gaussinit_gaussnoise}
    \end{subfigure}
    \hfill
    \begin{subfigure}{0.495\linewidth}
        \centering
        \includegraphics[width=1\linewidth]{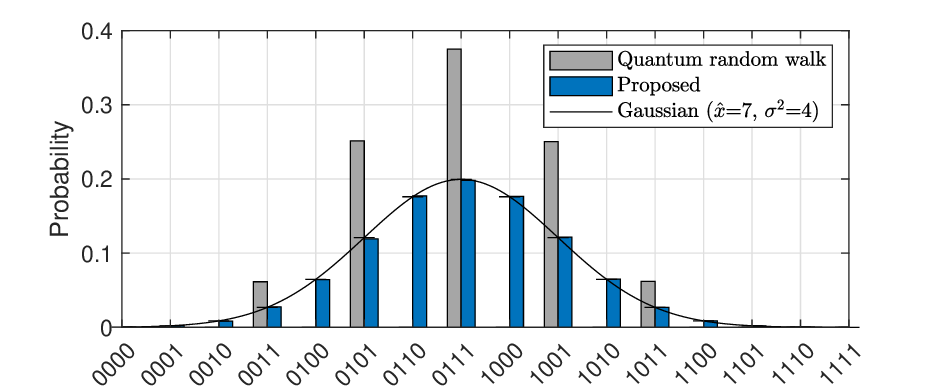}
        \vskip -0.5pc
        \caption{}
        \label{fig:impulseinit_gaussnoise}
    \end{subfigure}
    \hfill
    \begin{subfigure}{0.495\linewidth}
        \centering
        \includegraphics[width=1\linewidth]{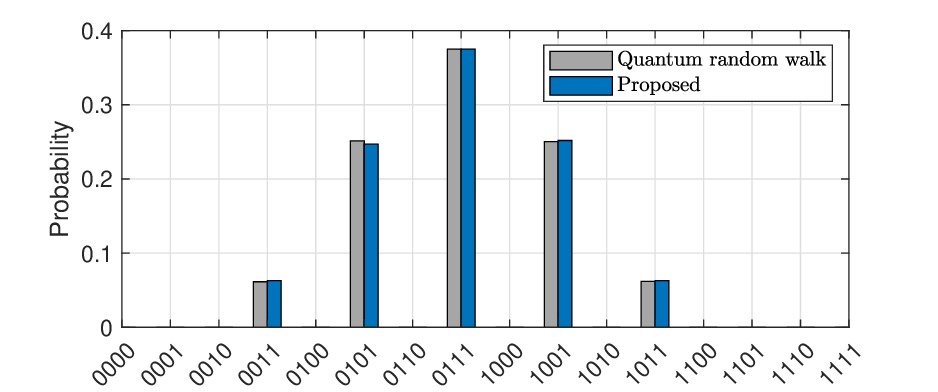}
        \vskip -0.5pc
        \caption{}
        \label{fig:impulseinit_randomwalk}
    \end{subfigure}
    \vskip -0.5pc
    \caption{Approximate probability densities obtained via quantum sampling for the four test cases, with solid curves indicating the exact densities. (a) Case 1. (b) Case 2. (c) Case 3. (d) Case 4}
    \label{fig:results}
\end{figure*}
Figure \ref{fig:results} shows the approximate probability distributions obtained through quantum sampling, with the solid curves indicating the exact density. As expected, the proposed method produces results that are very close to the exact distributions. From the results for Case 1 (Fig. \ref{fig:guassinit_1step}) and Case 2 (Fig.\ref{fig:gaussinit_gaussnoise}), we can observe that the accuracy of the QRW-based method improves as the number of repeats increases.
Table \ref{tab:complexity} summarizes the quantum resource requirements for each test case. 
Each circuit is decomposed into elementary operations to count single-qubit gates (1Q gates), two-qubit controlled-NOT gates (2Q gates), and overall circuit depth. Compared with the QRW-based method, the proposed QFT-based method requires approximately 5--20 times fewer gates and achieves circuit depths that are 8--31 times smaller. Since deeper circuits are more susceptible to errors and decoherence on current quantum hardware, reducing circuit depth is particularly important; current 
NISQ devices typically support effective depths below $10^3$ (\cite{PrYoBaBl:25}). For reference, the QRW-based circuits were implemented exactly as described in (\cite{Go:24a}), and their gate counts and depths may be reduced through transpilation, but the relative advantage of the proposed method remains unchanged.

\begin{table}[]
    \centering
    \caption{Gate counts and circuit depths for diffusion implementations}
    \label{tab:complexity}
    \renewcommand{\arraystretch}{1.25}
    \begin{tabular}{c c c r r r}
        \hline
        \textbf{Method} & \textbf{$p(\mathbf{x}^\mathrm{adv})$} & \textbf{\makecell{Steps/ \\$p(\mathbf{w})$}} & \textbf{\makecell{1Q \\ gates}} & \textbf{\makecell{2Q \\ gates}} & \textbf{Depth}\\
        \hline
        \multirow{3}{*}{\makecell{Quantum \\ random walk}}
            & $\mathcal{N}(7, 1^2)$ & 1 step        & 576  & 361  & 732  \\
            & $\mathcal{N}(7, 1^2)$ & 4 steps       & 2,259 & 1,411 & 2,832 \\
            & $P(7) = 1$  & 4 steps       & 2,258 & 1,411 & 2,832 \\
        \hline
        \multirow{4}{*}{\makecell{QFT-based \\ (proposed)}}
            & $\mathcal{N}(7, 1^2)$ & $\mathcal{N}(0, 1^2)$   & 104  & 66   & 89   \\
            & $\mathcal{N}(7, 1^2)$ & $\mathcal{N}(0, 2^2)$   & 104  & 66   & 89   \\
            & $P(7) = 1$  & $\mathcal{N}(0, 1^2)$   & 103  & 66   & 89   \\
            & $P(7) = 1$  & \makecell{4 steps\\(pseudo)} & 95  & 59   & 89   \\
        \hline
    \end{tabular}
\end{table}

\section{Conclusion}
\label{sec:5}
Quantum computing remains in its early stages of practical deployment, and many of its potential applications are still largely unexplored.
In this work, we presented a simple QFT-based quantum algorithm for implementing the diffusion step of grid-based filters.
Compared with existing coin-flip–style QRW implementations, the proposed approach achieves substantially shallower circuit depth and requires fewer quantum gates, while accurately accommodating a variety of probability distributions. 
It should be emphasized, however, that the proposed approach focuses specifically on the computation of probability densities and thus constitutes only one possible direction; depending on future advances in quantum hardware and algorithmic techniques, alternative formulations—including existing approaches—may still play an important role.

\section*{DECLARATION OF GENERATIVE AI AND AI-ASSISTED TECHNOLOGIES IN THE WRITING PROCESS}
During the preparation of this work the authors used ChatGPT in order to polish the language and enhance readability. 
After using this tool/service, the authors reviewed and edited the content as needed and take full responsibility for the content of the publication.

\bibliography{ifacconf}

@article{KuSaUy:21,
	title = {Survey on {Quantum} {Circuit} {Compilation} for {Noisy} {Intermediate}-{Scale} {Quantum} {Computers}: {Artificial} {Intelligence} to {Heuristics}},
	volume = {2},
	journal = {IEEE Transactions on Quantum Engineering},
	author = {Kusyk, Janusz and Saeed, Samah M. and Uyar, Muharrem Umit},
	year = {2021},
	pages = {2501616},
}

@article{Po:23,
	title = {Quantum {Bayesian} computation},
	volume = {39},
	issn = {1526-4025},
	number = {6},
	journal = {Applied Stochastic Models in Business and Industry},
	author = {Polson, Nick and Sokolov, Vadim and Xu, Jianeng},
	year = {2023},
	pages = {869--883},
}

@article{LaMa:23,
	title = {Quantum-enhanced {Markov} chain {Monte} {Carlo}},
	volume = {619},
	number = {7969},
	journal = {Nature},
	author = {Layden, David and Mazzola, Guglielmo and Mishmash, Ryan V. and Motta, Mario and Wocjan, Pawel and Kim, Jin-Sung and Sheldon, Sarah},
	month = jul,
	year = {2023},
	pages = {282--287},
}

@article{StUl:23,
	title = {Quantum {Computing} for {Applications} in {Data} {Fusion}},
	volume = {59},
	number = {2},
	journal = {IEEE Transactions on Aerospace and Electronic Systems},
    author = {{Stoo{\ss}}, Veit and Ulmke, Martin and Govaers, Felix},
	month = apr,
	year = {2023},
	pages = {2002--2012},
}

@inproceedings{McKl:22,
	title = {Multiple {Target} {Tracking} and {Filtering} {Using} {Bayesian} {Diabatic} {Quantum} {Annealing}},
	booktitle = {2022 {Sensor} {Data} {Fusion}: {Trends}, {Solutions}, {Applications} ({SDF})},
	author = {McCormick, Timothy M. and Klain, Zipporah and Herbert, Ian and Charles, Anthony M. and Angle, R. Blair and Osborn, Bryan R. and Streit, Roy L.},
	month = oct,
	year = {2022},
	pages = {1--9},
}

@inproceedings{NiMa:19,
	title = {Markov process simulation on a real quantum computer},
	volume = {2172},
	urldate = {2025-09-03},
	booktitle = {AIP Conference Proceedings},
	author = {Nikolov, Petar and Galabov, Vassil},
	year = {2019},
	pages = {090007},
}

@article{MaDuSt:25,
	author = {Matou\v{s}ek, J. and Dun\'{i}k, J. and Straka, O.},
	journal = {IEEE Signal Processing Magazine},
	title = {Lagrangian grid-based filters with application to terrain-aided navigation},
	year = {2025}}

@article{StDuTi:23,
	title = {Point-mass Filter with Functional Decomposition of Transient Density and Two-level Convolution},
	journal = {IFAC-PapersOnLine},
	volume = {56},
	number = {2},
	pages = {6934-6939},
	year = {2023},
	note = {22nd IFAC World Congress},
	issn = {2405-8963},
	author = {Ondřej Straka and Jindřich Duník and Petr Tichavský}
}

@ARTICLE{TiStDu:23,
	author    = {Petr Tichavsk\'{y} and Ond\v{r}ej Straka and Jind\v{r}ich Dun\'{\i}k},
	journal   = {IEEE Transactions on Signal Processing}, 
	title     = {Grid-Based {B}ayesian Filters With Functional Decomposition of Transient Density}, 
	year      = {2023},
	volume    = {71},
	number    = {},
	pages     = {92--104},
}

@article{DuStMaBl:22,
	author = {J. Dun{\'\i}k and O. Straka and J. Matou{\v s}ek and E. Blasch},
	journal = {Signal Processing},
	title = {Copula-based convolution for fast point-mass prediction},
	volume = {192},
	year = {2022}
}

@INPROCEEDINGS{AjSt:23,
	author    = {Ji\v{r}\'{\i} Ajgl and Ond\v{r}ej Straka},
	booktitle = {2023 26th International Conference on Information Fusion (FUSION)}, 
	title     = {Approximate fusion of probability density functions using {G}aussian copulas}, 
	year      = {2023},
	pages     = {1--7},
}

@article{MaBrDuPu:24,
  title     = {Tensor Train Discrete Grid-Based Filters: Breaking the Curse of Dimensionality},
  journal   = {IFAC-PapersOnLine},
  volume    = {58},
  number    = {15},
  pages     = {19--24},
  year      = {2024},
  note      = {20th IFAC Symposium on System Identification SYSID 2024},
  issn      = {2405-8963},
  author    = {J. Matou\v{s}ek and M. Brandner and J. Dun\'{\i}k and I. Pun\v{c}och\'a\v{r}},
}

@INPROCEEDINGS{Go:18,
	author={Govaers, Felix},
	booktitle={2018 21st International Conference on Information Fusion (FUSION)}, 
	title={On Canonical Polyadic Decomposition of Non-Linear Gaussian Likelihood Functions}, 
	year={2018},
	pages={1-7}
 }

@INPROCEEDINGS{PaZh:11,
	author={Pace, Michele and Zhang, Huilong},
	booktitle={14th International Conference on Information Fusion}, 
	title={Grid based PHD filtering by Fast Fourier Transform}, 
	year={2011},
	volume={},
	number={},
	pages={1-8},
}

@inproceedings{Go:24a,
	title = {A {Quantum} {Algorithm} for the {Prediction} {Step} of a {Bayesian} {Recursion}},
	booktitle = {2024 27th {International} {Conference} on {Information} {Fusion} ({FUSION})},
	author = {Govaers, Felix},
	month = jul,
	year = {2024},
	pages = {1--8},
}

@inproceedings{Go:24b,
	title = {On a {Quantum} {Realization} of the {Bayesian} {Filtering} {Using} the {Log}-{Homotopy} {Flow}},
	booktitle = {2024 {IEEE} {International} {Conference} on {Multisensor} {Fusion} and {Integration} for {Intelligent} {Systems} ({MFI})},
	author = {Govaers, Felix},
	month = sep,
	year = {2024},
	pages = {1--6},
}

@inproceedings{DaHuNo:10,
author = {Daum, Fred and Huang, Jim and Noushin, Arjang},
booktitle = {Proceedings in SPIE: Signal Processing, Sensor Fusion, and Target Recognition XIX},
month = apr,
title = {{Exact particle flow for nonlinear filters}},
volume = {7697},
address = {Orlando, FL},
year = {2010}
}

@article{Ke:03,
	title = {Quantum random walks: {An} introductory overview},
	volume = {44},
	number = {4},
	journal = {Contemporary Physics},
	author = {Kempe, J},
	month = jul,
	year = {2003},
	pages = {307--327},
}

@misc{Dr:00,
	title = {Addition on a {Quantum} {Computer}},
	publisher = {arXiv},
	author = {Draper, Thomas G.},
	month = aug,
	year = {2000},
}

@article{RuGaJc:17,
	title = {Quantum arithmetic with the quantum {Fourier} transform},
	volume = {16},
	number = {6},
	journal = {Quantum Information Processing},
	author = {Ruiz-Perez, Lidia and Garcia-Escartin, Juan Carlos},
	month = jun,
	year = {2017},
}

@article{PrYoBaBl:25,
	title = {Benchmarking quantum computers},
	volume = {7},
	issn = {2522-5820},
	number = {2},
	journal = {Nature Reviews Physics},
	author = {Proctor, Timothy and Young, Kevin and Baczewski, Andrew D. and Blume-Kohout, Robin},
	month = feb,
	year = {2025},
	pages = {105--118},
}

\end{document}